\documentclass{aastex}
\usepackage{spr-astr-addons}
\usepackage{url}
\usepackage{rotating}

\RequirePackage{color}

\begin{document}

\title{Study of the inner disk of the Herbig star MWC480}
\slugcomment{Not to appear in Nonlearned J., 45.}
\shorttitle{Short article title}
\shortauthors{Autors et al.}

\author{N. Jamialahmadi\altaffilmark{1}} 
\affil{Laboratoire D.-L. Lagrange, UMR 7293 UNS-CNRS-OCA, Boulevard de l'Observatoire, CS 34229, 06304 NICE Cedex 4, France}
\and \author{B. Lopez\altaffilmark{1}}
\and
\author{Ph. Berio\altaffilmark{1}}
\and
\author{S. Flament\altaffilmark{1}}
\and
\author{A. Spang\altaffilmark{1}}


\altaffiltext{1}{Laboratoire D.-L. Lagrange, UMR 7293 UNS-CNRS-OCA, Boulevard de l'Observatoire, CS 34229, 06304 NICE Cedex 4, France.}

\begin{abstract}
The inner structure and properties (temperature, mass) of the circumstellar disk of Herbig star MWC480 are studied by stellar interferometry method used in the infrared and are interpreted using semi-analytical models. From these models, the SED (Spectral Energy Distribution) was fitted and multi-wavelength intensity map of the source were calculated. The intensity map provides the input for modeling the Keck Interferometer (KI) data in the near-infrared (near-IR) and the data of the Very Large Telescope Interferometer (VLTI) with the mid-infrared instrument MIDI. We conclude that with our limited set of data, we can fit the SED, the Keck visibilities and the  MIDI visibilities using a two-components disk model. Furthermore, we suspect that MWC480 has a transitional dusty disk. However, we need more MIDI observations with different baseline orientations to confirm our modeling.
\end{abstract}

\keywords{circumstellar dust --
                planetary system --
                star: MWC480 -- techniques: interferometric  }


\section{Introduction}
Circumstellar disks around young stellar objects provide the physical conditions at the origin of the formation of planets. A new class of objects has been identified using mid- and far-infrared telescopes such as Spitzer, the pre-transitional and transitional disks \citep[e.g.,][]{2010ApJ...717..441E}. Pre-transitional disks have a typical spectral energy distribution (SED) with a near-infrared excess resulting from the emission of hot dust and gas located in an inner disk, a dip in the mid-infrared range likely caused by a gap, and, at longer wavelengths, the signature of an optically thick outer disk.

MWC480 (HD 31648, ${A2/3ep+sh}$) is an Herbig Ae star of 1.8 $ M_{\odot} $ \citep{2000ApJ...545.1034S} located at d=137$\pm$31 pc \citep{2007A&A...474..653V}. This star is one of the brightest Herbig Ae stars at millimeter wavelengths \citep{1997Natur.388..555M} surrounded by a Keplerian disk \citep{1997ApJ...490..792M,2000ApJ...545.1034S, 2007A&A...467..163P}. These authors have mapped the thermal dust millimeter-continuum and gaseous CO emission towards MWC480 and found that a circumstellar (CS) disk surrounding the star has an extent of 85 AU (FWHM) and an inclination angle i $\sim$ $30^{\circ}$. The disk continuum emission was resolved by \citet{2006A&A...460L..43P}. The IR excess of the object was investigated by \citet{1981ApJ...247.1024S}, \citet{2001A&A...365..476M}. The gas disk of MWC480 can be studied in spectral lines since hydrogen in accretion flows or in the innermost regions of outflows can be ionized. One of the hydrogen spectral lines which has special importance is the Br$\gamma$ line, that is strongly correlated with accretion onto young stars \citep{1998AJ....116.2965M}. 

 In this paper, we report the first interferometric observations of MWC480 using the VLTI instrument MIDI \citep{2003Ap&SS.286...73L}, observing in the N band (8--13 $\mu$m). We simultaneously modeled the SED, the near-IR and mid-IR interferometric data of MWC480 to constrain the overall spatial structure of the inner disk region.

Section 2 summarizes the observations and data processing. Section 3 describes our modeling approach, making use of semi-analythical models of one-component and two-components disk and the results. Section 4 summarizes our work and outlines some perspectives.

\section{Observations}
 \subsection{MIDI observations and data reduction}
   \subsubsection{Observations}
Using the instrument MIDI of the VLTI \citep{2003Ap&SS.286...73L} in operation at the ESO Paranal Observatory, MWC480 has been observed during one night in 2007. This observation was carried out on the 4th of February 2007 in one run using the 8m Unit Telescopes (UTs).
The observations were performed using the prism as dispersive element giving a spectral resolution of R $\sim$ 30 in the N-band. 
We obtained three fringe and photometry measurements for MWC480 using the HIGH-SENS mode. In this mode, the photometry or total flux is measured just after the fringe acquisition.  
In order to calibrate the visibility of the science target, we used HD20644 as calibrator. This calibrator was selected from the MIDI calibrator list with the SearchCal  JMMC$\footnote{http://www.mariotti.fr/}$ tool. This tool provides a validated database for the calibration of long-baseline interferometric observations \citep{2006A&A...456..789B}.

MWC480 also was observed with the KI in 2007 by \citet{2009ApJ...692..309E}. We downloaded the reduced data from the Keck Archive (See. Fig.1 [Bottom-left]). The peak in the Keck data, which has higher visibility value is related to the Br$\gamma$ emission at 2.165$\mu$m. 

We show a summary of the observing log, containing the length and Position Angle (P.A.) of the projected baselines in Table \ref{table:1}. 

The UV coverage of the interferometric observations is shown in Fig.1 [Top-left]. 

\subsubsection{Data reduction}
The calibration of the visibility measurements and total flux of MWC480 were performed using the data-reduction software package named MIDI Interactive Analysis (MIA) and Expert WorkStation $\footnote{The software package is available at http://home.strw.leidenuniv.nl/$\sim$jaffe/ews/index.html the software manual is available at http://home.strw.leidenuniv.nl/$\sim$jaffe/ews/MIA+EWS-Manual/index.html}$(EWS). This software performes a coherent analysis of dispersed fringes to estimate the complex visibility of the source. The method and the different processing steps are described in \cite{2004SPIE.5491..715J}. The calibrated visibilities were then obtained by dividing each raw visibility measurement by the instrumental visibility measured on the closest calibrator in time. 
 
 \subsection{Spectroscopic observations} 

Since the MIDI observations have been done in HIGH-SENCE mode in 2007, we tried to obtain the uncorrelated flux using VLTI/MIDI for wavelengths 8--13.5 $ \mu $m. In Fig.1 [Bottom-right], we show this flux in the SED with pink color.
To have the same date observations with the intereferometric observations in 2007 in the SED, we used the SpeX data for wavelengths 0.8--5.2 $ \mu $m and the BASS data for wavelengths 5.4--14 $ \mu $m {\bf{\citep{2012ApJ...753..153K}}}. We show these data in the SED with green color (see Fig.1 [Bottom-right]). 

\subsubsection{Variability} 
 
\citet{2008ApJ...678.1070S}, \cite{2010ApJ...719.1565G} and \citet{2012ApJ...753..153K} figured out the variability of MWC480 in the near- and mid-IR emission. It is now well-established that the near-IR and mid- to far-IR variability  are often anti-correlated, at least in transitional disks \citep{2012ApJ...753..153K}. The most likely scenario is changes in the scale height of the inner disk. This leads to changes in the shadowing of the outer disk so that the illumination by the central star changes with time. This affects both the scattered light by dust particles \citep[e.g.,][]{2012ApJ...753..153K} and thermal emission \citep[e.g.,][]{2007ApJ...661..374T}. \citet{2012ApJ...753..153K} do show that this variability in the near-IR is due to scale height variability of the dust disk at the sublimation radius, which can affect the shadowing of entire disk. The detection of mid-IR photometric variability prompted Sitko et al. (2008) to suggest that the disk of MWC480 might be variably illuminated, with scattered-light imagery showing apparently variable disk structure. As we mentioned above, in most case of transitional disks  the near- and mid- to far-IR variability are anti-correlated. For instance, if the scale height of the dust disk at the sublimation radius is smaller than usual, which means that the object is observed in its minimum brightness state in the near-IR, the outer disk is expected to be detected more in scattered light than in its maximum brightness state \citep{2012ApJ...753..153K}. According to Fig.1 [Bottom-right], for the SED, we used the ISO data obtained in 1998 for wavelengths 2.3--198 $ \mu $m \citep{2002A&A...385..546C}. The ISO data have almost less than 10$ \% $ difference with the brightness state in 2007 data in the near-IR according to Fig. 5 of \citet{2012ApJ...753..153K}. We used also the Spitzer data for wavelengths of 3--198 $ \mu $m by {\bf{Houck et al. (2004)}} \citet{2004SPIE.5487...62H}. These data were obtained when the star was in the maximum brightness state in the near-IR.

\paragraph{Results} 
According to Fig.1 [Bottom-right], the total flux obtained by the MIDI observation in 2007 between wavelengths 8--13.5 $ \mu $m are almost consistent with the BASS data in 2007. Since there were no data available for wavelengths 20--200 $ \mu $m in 2007, we used the ISO data from 1998, which have less than 10$ \% $ difference with the brightness state in the near-IR with 2007 data. As we mentioned above, the Spitzer data are in the maximum brightness state in the near-IR of the SED and roughly 2 factor above the 2007 data. We over plotted these data just to show how the inner rim scale height differences can affect the shadowing of the outer disk compared to 2007 and 1998 data.
As we represent in Fig. 1 [Bottom-right], although the Spitzer data are a factor of almost 2 times above the 1998 and 2007 data in the near-IR, for wavelengths 10--200 $ \mu $m they are a factor of roughly 2 below the 2007 and 1998 data. 
So, the changing in the scale height of the inner disk leads to changes in the shadowing of the outer disk so that the illumination by the central star changes with time. This affects both the scattered light \citet{2012ApJ...753..153K} and thermal emission (e.g., \cite{2007ApJ...661..374T}).

\begin{table*}
\caption{Log of observations. The MIDI observations come from a program prepared by Di Folco et al. (2007), the Keck observations come from Keck Archive with program ID of 'ENG' by \citet{2007ApJ...669.1072E}.}
\label{table:1} 
\centering
\begin{tabular}{l| c |l| c |l c  c  c}
\hline\hline
 \textbf{Instrument}& \textbf{Telescopes} & \textbf{Date} & \textbf{$B_{p}$ [m]}& \textbf{P.A. [$^{ \circ }$]}\\
\hline
MIDI & UT2-UT3& 2007--02--04&42.8& $52^{\circ }$\\
HIRES& K1-K2 &2007--07--03  & 84.9 & $48^{\circ }$\\

\hline\hline
\end{tabular}
\end{table*}

%

 \begin{figure*}[htb]
  \begin{center}
  \includegraphics[width=7cm, height=5cm]{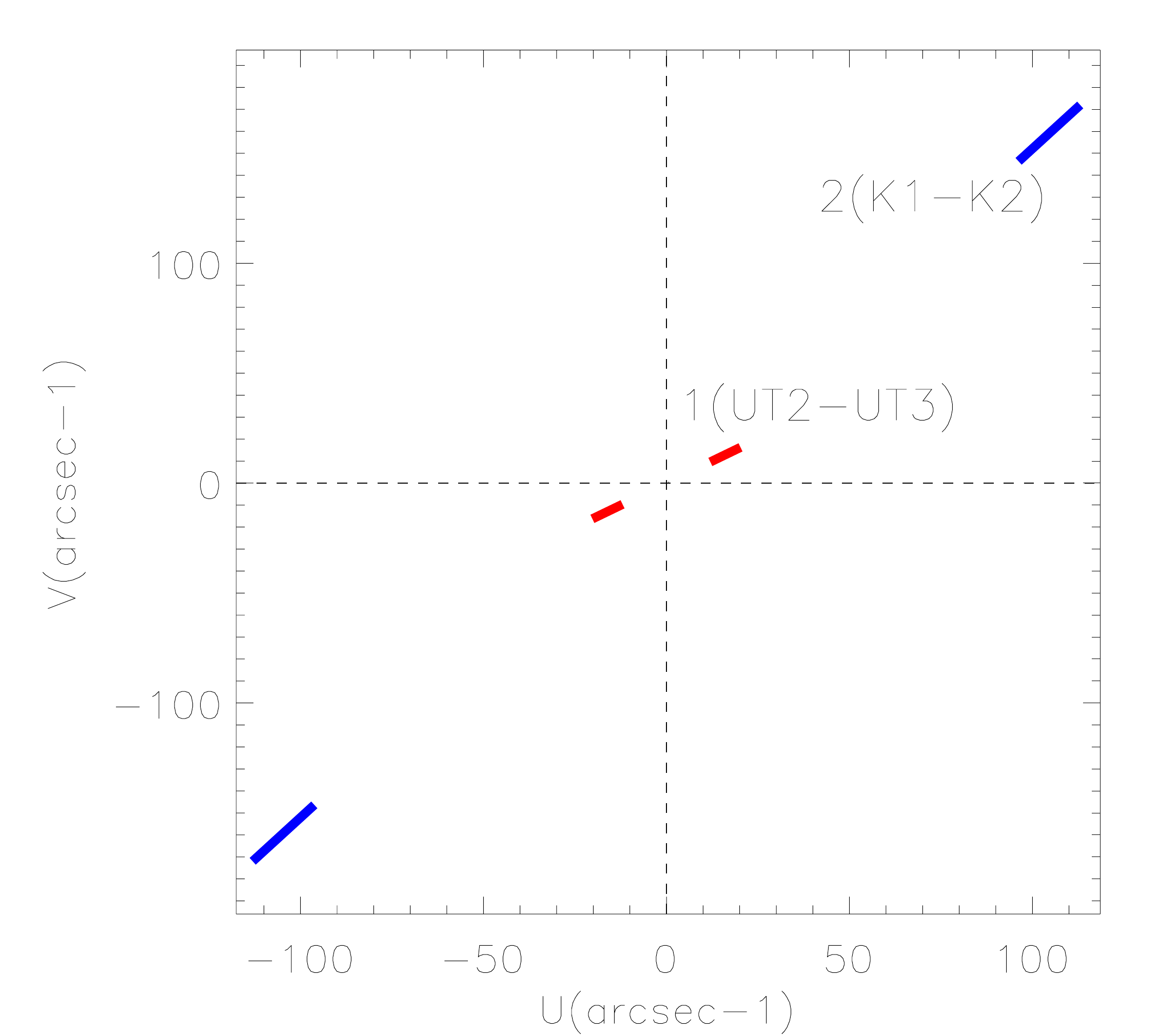}
   \includegraphics[width=7cm, height=5cm]{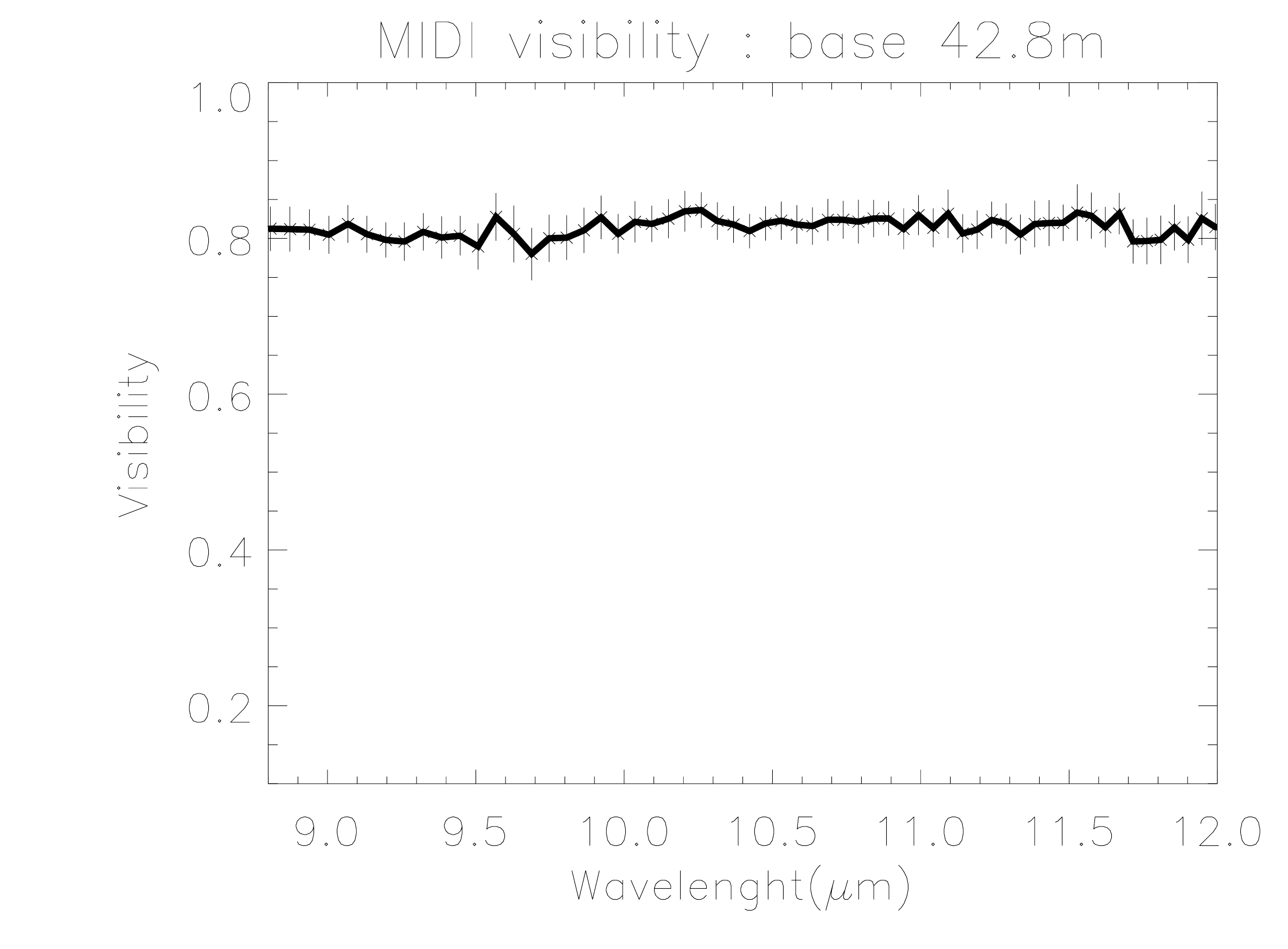} \\
   \includegraphics[width=7cm, height=5cm]{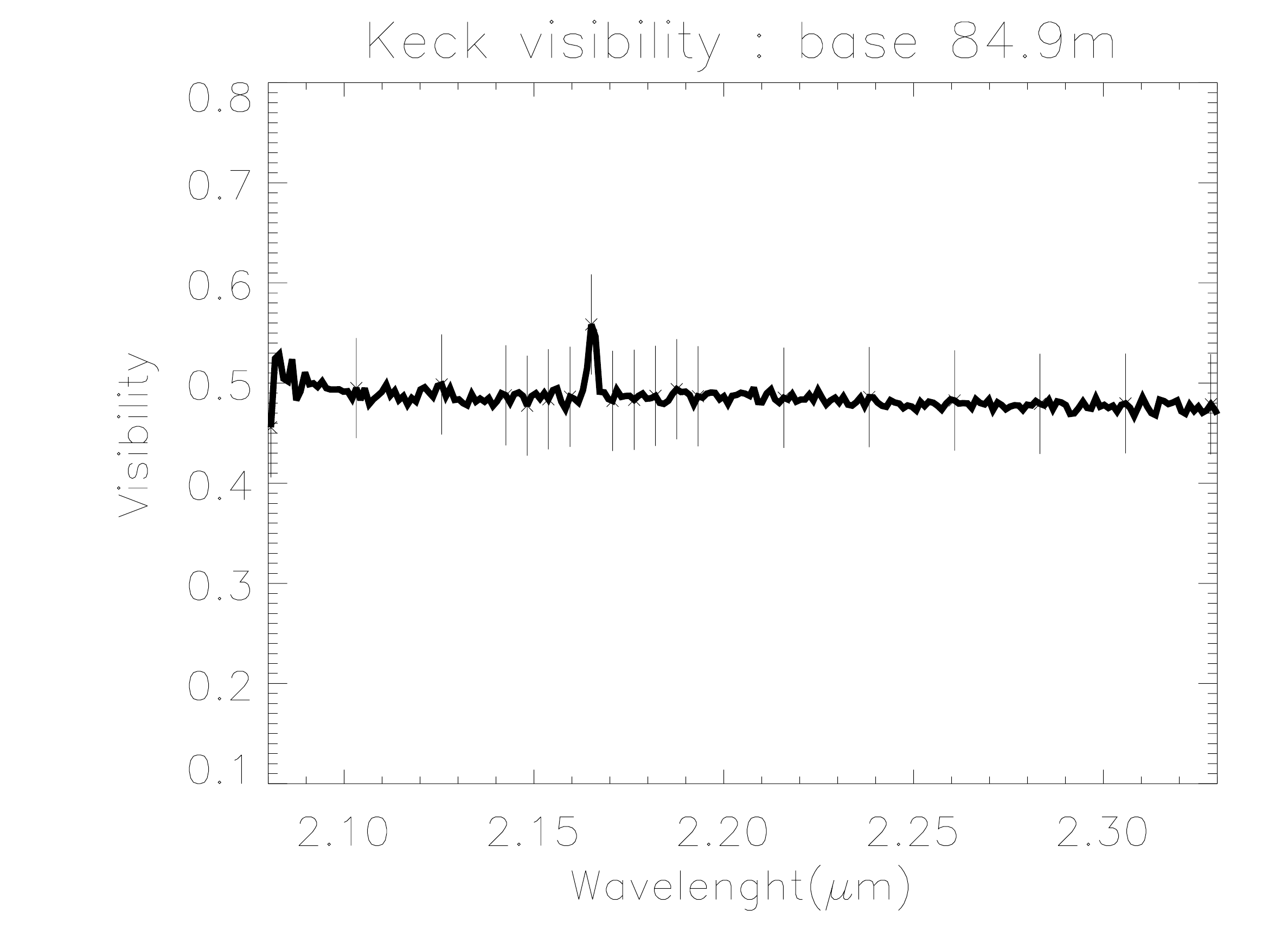}
   \includegraphics[width=7cm, height=5.5cm]{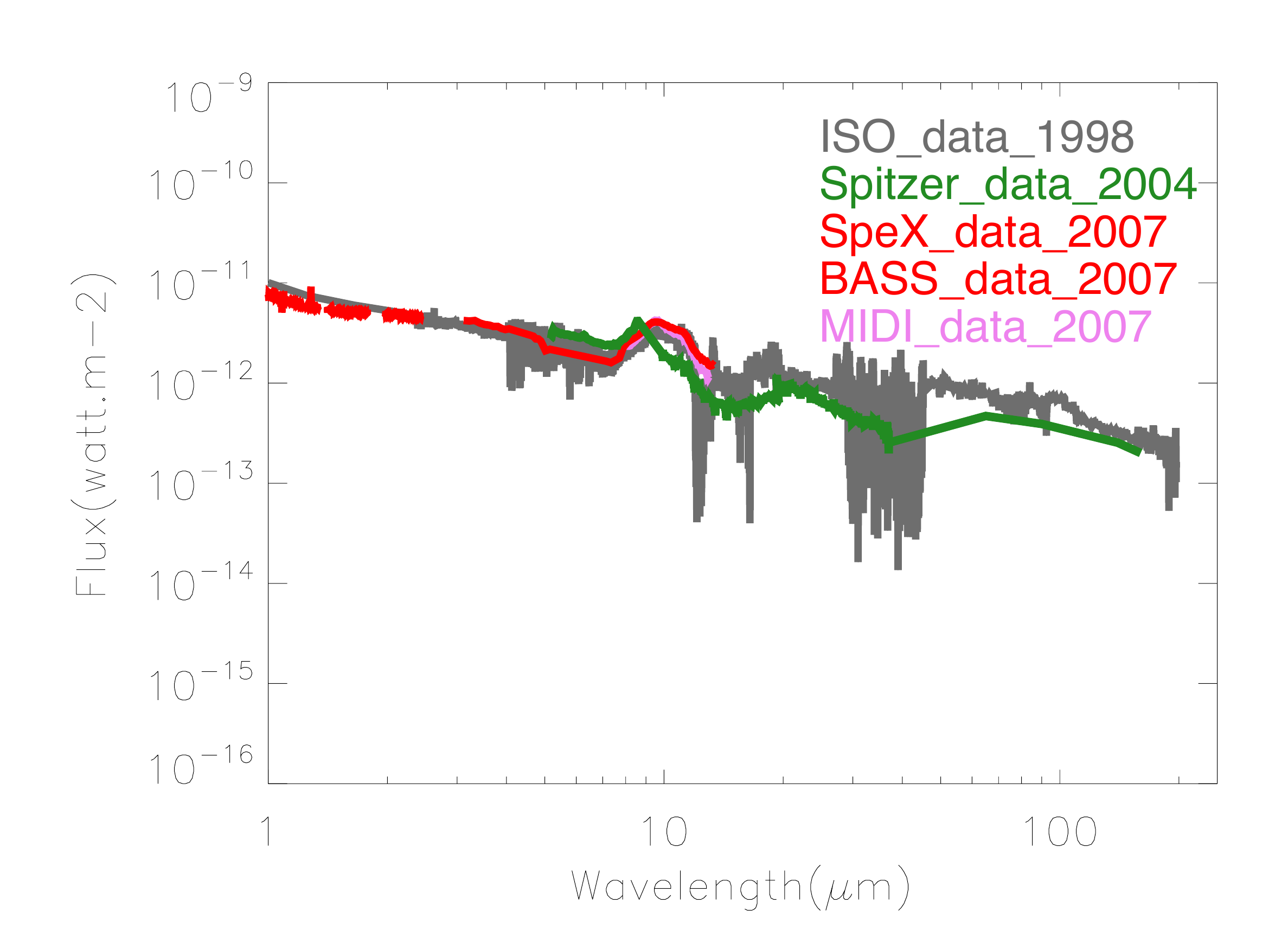}
  \end{center}
  \caption{Top left: UV coverage for the two sets of observations; the red color refers to B =42.8 m and blue color refers to B =84.9 m (as detailed in Table 1). Top right: Measured N-band visibility of MWC480 (with error bars) as a function of wavelength. Bottom left: Measured K-band visibility of MWC480 (with error bars, which are selected for few points because of plenty of Keck data). Bottom right: ISO data in 1998 between wavelengths 2.2 $ \mu $m and 200 $ \mu$m in gray color, averaged MIDI uncorrelated spectrum in 2007 between wavelengths 8 $ \mu $m--13.5 $ \mu $m in pink color, SpeX and BASS data in 2007 for wavelengths 0.8 $ \mu $m--5.5 $ \mu $m and 5.8 $ \mu $m--14 $ \mu $m respectively in red color and Spitzer data in 2004 for wavelengths 4 $ \mu $m--180 $ \mu $m in green color.}
  \label{Lable}
 \end{figure*}

\section{Modeling}

We aim to develop and compare some semi-analytical models to study the circumstellar environment of Herbig star MWC480. The semi-analythical models simulated the disk emission assuming different possible geometries and dust opacities. The models allow us to reproduce the KI and MIDI visibilities and the SED. We explored a range of possible parameters describing the object and including the size and mass of the disk for example.

%
   

   We constructed a disk model with a temperature gradient. The temperature decreases with increasing radius. In our model, we computed the visibilities from the Fourier transforms of the image for the star plus the disk at each wavelength. 
The star was modeled by a blackbody $B_{\lambda}( T_{\star} )$. $B_{\lambda}$ is the monochromatic intensity represented by the planck function. 
The star has an angular radius of $\alpha_{\star}$=$R_{\star}$/d in radian, where d is the distance of the star and $R_{\star}$ is the stellar radius. 

The total received flux is $B_{\lambda}( T_{\star} )$ $\times$ ($\pi$ ${\alpha_{\star}}^{2}$) expressed in W${m^{-2}}$ ${m^{-1}}$, where $\pi$ ${\alpha_{\star}}^{2}$ is the solid angle of the star seen from the observer and $T_ {\star}$ is the stellar temperature.

The disk was also modeled by a black body emission $B_{\lambda}[T(r)]$, where $T(r)$ is the temperature law according to the distance r to the star. In fact for an optically thick disk, the disk will emit as a black body. However, as described below, the vertical optical depth of the disk itself is related to the opacity of the dust material that has to be taken in to account.
We adopted a power law form for $T(r)$:
   
\begin{equation}
      T(r) =T_ {in} \left(\frac{r}{r_{in}}\right)^{-q}\,,
       \end{equation} 
  
%

 with q ranging from 0.5 (flared irradiated disks) to 0.75 (standard viscous disk or flat irradiated disks), see e.g., \citet{1981ARA&A..19..137P}.
 $T_{in}$=$T_ {g}$ is the temperature of a free grain$\footnote{A grain illuminated directly by the central star assuming no radiative exchange with other grains.}$ located at $r$=$r_{in}$ which is the inner radius of the disk.  
 
Equating the absorption and the emission of a grain with non-chromatic (gray) absorptivity,  one can calculate the $T_{in}$ at $r_{in}$ in Eq.(1):
   \begin{equation}
      T_{in}=T_ {g} =T_ {\star} \left(\frac{R_{\star}}{2r_{in}}\right)^{\frac{1}{2}}\,,
   \end{equation}

when the disk is optically thick (we are interested in the optical depth in the vertical direction, assuming the observer is looking nearly perpendicularly to the disk), then for each surface area $s_{disk}$ of the disk, the observer receives $B_{\lambda}[T(r)]$ $\times$ ($s_{disk}/{d^{2}}$). The quantity $s_{disk}/{d^{2}}$ represents the solid angle of each elementary surface area$\footnote{The elementary surface area of the disk is defined by our pixel size in the images of our model.}$ of the disk seen at the distance d
Defining  $r_{in}$, $r_{out}$, $T(r)$, $\Sigma(r)$ and $\kappa_{\lambda}$ completely characterizes the disk. $r_{out}$ is the outer radius of the disk, $\kappa_{\lambda}$ is the opacity of the dust and $\Sigma(r)$ is the surface density law. The quantity $B_{\lambda}( T_{r} )[1-exp(-\tau_{\lambda}(r))]$ represents the general expression of the brightness of the surface of the  disk according to  $\tau_{\lambda}(r)$, the optical depth in the vertical direction \citep{1990AJ.....99..924B,2001ApJ...560..957D}. In this case, the observer receives $B_{\lambda}[T(r)]$  $[1-exp(-\tau_{\lambda}(r))]$ ($s_{disk}/{d^{2}}$). When the disk is inclined by an angle i, this quantity becomes
\begin{equation}
       B_{\lambda} \left[  T(r)\right] \left[1-exp\left(-\frac{\tau_{\lambda}(r)}{cos(i)}\right)\right]  \left(\frac{s_{disk}}{d^{2}}\right)
   \end{equation}

 The relation of $\tau_{\lambda}(r)$ with surface density and $\kappa_{\lambda}$ is described as below
\begin{equation}
       \tau_{\lambda}(r) ={\Sigma(r) \kappa_{\lambda}}
   \end{equation} 

We used the dust opacity from Fig. 3 of \citet{2010MNRAS.406.1409T}, computed from Mie theory for a grain size distribution following the "${a}^{-3.5}$" power law and with a minimum size of 0.02 $\mu$m and three values of the maximum grain size ($a_{max}$=[10, 50, 200] $\mu$m). 
The power law form for $\Sigma(r)$ is described as

      \begin{equation}
      \Sigma(r) =\Sigma_ {0} \left(\frac{r}{r_{0}}\right)^{-p}\
   \end{equation}
 In the models of \citet{2001ApJ...560..957D} and \citet{1997ApJ...490..368C}, $\Sigma_ {0}$ has been considered at r=1 AU, while in e.g., \citet{1998A&A...338L..63D}, \citet{2011A&A...535A.104D}, who used millimeter observations, $\Sigma_ {0}$ is assumed to be defined from the outer radius of the disk. Since we measured directly by long baseline interferometry at 2 $\mu$m the warm dust located at the inner radius, we choosed to consider $\Sigma_ {0}$ from ${r_{in}}$. Therefore in our model,  the $\Sigma_ {0}=\Sigma_ {in}$ and ${r_{0}}$=${r_{in}}$ gives
   
    \begin{equation}
      \Sigma(r) =\Sigma_ {in} \left(\frac{r}{r_{in}}\right)^{-p}\,,
   \end{equation}  
     where p ranges from 1 (assuming constant mass accretion rate at constant viscosity) to p= 1.5 as inferred for the MMSN (Minimum Mass Solar Nebula) \citep{1997LPI....28.1517W} and assumed often as a basis in other disk models (e.g., \citet{1997ApJ...490..368C}; \citet{2001ApJ...560..957D}; \citet{2009ApJ...692..309E}). $\Sigma_ {in}$ is related to the total mass amount of the dust. The mass of the dust is given by
  \begin{equation}
       M_{dust}=\int_0^{2\pi} \int_{r_{in}}^{r_{out}} \mathrm\Sigma(r) \,\mathrm{d}r \mathrm{d}\theta\
   \end{equation} 
 Combining Eq.(6) and Eq.(7) gives
  \begin{equation}
       \Sigma_{in}=\frac{M_{dust}}{2\pi {r}_{in}^{2} f}\,,
   \end{equation} 
   where 
   \begin{equation}
   f=\frac{1}{2-p} \left[\left(\frac{r_{out}}{r_{in}}\right)^{2-p}-1\right]
   \end{equation} 
    Combining Eq.(4) and Eq.(6) gives

\begin{equation}
       \tau_{\lambda}(r) =\tau_ {\lambda,in} \left(\frac{r}{r_{in}}\right)^{-p}\,,
   \end{equation}  
   where
   
  \begin{equation}
       \tau_ {\lambda,in} =\Sigma_ {in} \kappa_{\lambda}\
   \end{equation}  

\begin{table*}
\caption{The best parameters and their explored ranges for the one-component disk model.}
\label{table:2}  
\centering
\begin{tabular}{c | c | c | c | c | c | c}
\hline\hline
\textbf{Parameters} & \textbf{Best-Values} & \textbf{Explored Ranges}& \textbf{$ \chi_{r \hspace{0.1cm} SED}^{2} $} &\textbf{ $ \chi_{r \hspace{0.1cm} vis1}^{2} $}& \textbf{$ \chi_{r \hspace{0.1cm} vis2}^{2} $}&\textbf{$ \chi_{r \hspace{0.1cm} total}^{2} $} \\ [0.5ex] 
\hline
\textbf{$M_{dust}$ ($a_{max}$=10 $\mu$m)}&2.4  $\times$ ${{10}^{-11}}$ ${M_{\odot}}$ & ${10}^{-12}$...${10}^{-7}$ ${M_{\odot}}$& & & &   \\
\textbf{$M_{dust}$ ($a_{max}$=50 $\mu$m)}&0.5  $\times$ ${{10}^{-10}}$ ${M_{\odot}}$ & ${10}^{-12}$...${10}^{-7}$ ${M_{\odot}}$& & & &   \\
\textbf{$M_{dust}$ ($a_{max}$=200 $\mu$m)}&${{10}^{-10}}$ ${M_{\odot}}$ & ${10}^{-12}$...${10}^{-7}$ ${M_{\odot}}$ & 36.08& 0.03& 2.15& 12.75\\
\textbf{p}&${1.5}$&0.1...1.98 & & & & \\
\textbf{q}&${0.5}$&0.4...0.9 & & & & \\
\textbf{$r_{in}$}& ${0.27}$ AU&0.1...0.4 AU & & & &  \\
\textbf{$r_{out}$}& ${80}$ AU & fixed & & & & \\[1ex]

\hline\hline
\end{tabular}
\end{table*}
We modeled circumstellar Br$\gamma$ emission by including additional flux at 2.165$\mu$m produced in the inner region of the disk as an optically thin isothermal gaseous disk. The flux received from Br$\gamma$ emission is calculated as 

\begin{equation}
B_{\lambda}\left[T(r)\right] \left[1-exp\left(-\frac{\tau_{\lambda}(r)}{cos(i)}\right)\right] \left(\frac{s_{ring}}{d^{2}}\right) , 
\end{equation}

where $s_{ring}/{d^{2}}$ represents the solid angle of each surface area of the ring of Br$\gamma$ emission seen at the distance d.

To minimize the parameters of our model, we used the parameters of \citet{2009ApJ...692..309E}  in order to reproduce the effect of Br$\gamma$ in the Keck visibility. Then we assumed that the emission arises from a gaseous ring with $\tau_{\lambda,in}$  $\approx $ 0.08 and $T(r_{in})$ $\approx $ 3500K located in ${r_{in}}$ $\approx $ 0.07 AU. In our model, we assumed that the outer radius of the gaseous ring is 0.09 AU, while in \citet{2009ApJ...692..309E}  the gaseous disk was extended until the sublimation radius. In fact, Br$\gamma$ emission may even arise from a more compact region, but we could not constrain such small size scales with the available angular resolution and the limitation of our model.

\subsection{Application to the one-component disk model}
 The excess in the infrared for MWC480 has been modeled by \citet{2007ApJ...669.1072E}. However, after considering a thick disk emission, a residual near-IR excess between $\sim$ 2--10 $\mu$m could not be explained. \citet{2009ApJ...692..309E} then considered a gaseous disk plus a shell of dust to reproduce the whole infrared excess including the near-IR. We rather used an optically thin disk considering the temperature and surface density law according to Eq. (1) and Eq.(6).

For the SED, we considered only the SpeX and BASS data in 2007 for wavelengths 0.8--14 $\mu$m and ISO data in 1998 for wavelengths 15--200 $\mu$m and an over plot of total flux of MIDI data in 2007 for wavelengths 8--13.5 $\mu$m. In this model, the outer radius was fixed to a value of 80 AU \citep{1997ApJ...490..792M}. To minimize the number of free parameters, the inclination and P.A. of the disk were fixed. We fixed the inclination of the disk to the value derived from \citet{2000ApJ...545.1034S}, \citet{2006A&A...460L..43P}, \citet{2012A&A...537A..60C}, namely i=$37^{ \circ}$. The position angle was fixed to the average disk semi-minor axis P.A. of $57^{ \circ}$ derived from millimeter interferometry for an average of 5 CO and HCO+ transitions \citep{2007A&A...467..163P} and the P.A. of the jet obtained by \cite{2010ApJ...719.1565G}.
In this model, we have four free parameters:

\begin{enumerate}
\item[1)] The dust mass for each maximum grain size $M_{dust}$,
\item[2)] the surface density power-law exponent p,
\item[3)] the temperature power-law exponent q,
\item[4)] the inner radius $r_{in}$,  
\end{enumerate}

\paragraph*{-} The effect of mass of the dust species:

 The first important free parameter in our model is the total mass of the dust disk, $M_{dust}$. For three maximum sizes of grains in a size grain distribution, $a_{max}$=10 $\mu$m, $a_{max}$=50 $\mu$m and $a_{max}$=200 $\mu$m, we explored the total mass of the dust in the range $\approx $ ${10}^{-12}$--${10}^{-7}$ ${M_{\odot}}$. We derived the best value for each maximum sizes of grains (see Table 2). Increasing the mass of the dust further with respect to $a_{max}$, is not consistent with the SED in the range 2 $\mu$m to 10 $\mu$m, because larger masses cause an excess in the near-IR. But it can reproduce the flux of the SED in the range 10 $\mu$m to 200 $\mu$m.
 On the other hand, increasing the mass of the dust generates a decrease of the Keck and MIDI visibilities. Our model is formed by an unresolved star and a resolved disk surrounding the star. Increasing the dust mass, increases thus the flux ratio of the disk and star. Therefor, the visibility value decreases and disk becomes more resolved.

 \paragraph*{-} The effect of the exponent of the surface density law, p:
  
 The exponent of the surface density law, p, for a constant dust disk mass respect to the maximum grain size (see. Table 2) was explored in the range 1-2. The best value we found, was 1.5 as inferred for the protosolar nebula by \citet{1997LPI....28.1517W}. Increasing this exponent, meaning that the dust surface density decreases with a steeper slope, increases the flux of the SED just for the wavelengths from 1 $\mu$m to 13 $\mu$m. But it does not have any strong effect at longer wavelengths. Besides, it decreases the Keck visibilities and increases the MIDI visibilities. Decreasing this exponent down to p=1,  decreases the flux of the SED from 1 $\mu$m to 13 $\mu$m and the MIDI visibilities and increases the Keck visibility.

 \paragraph*{-} The effect of the exponent of the temperature law, q:
 
We have tested this exponent in the range 0.4-0.9. We derived the best value of q=0.5. Increasing the exponent causes decreasing the flux in the broad band SED especially at longer wavelengths. The Keck and MIDI visibilities both increase.

  \paragraph*{-} The effect of the inner radius of the disk, $r_{in}$: 
 
 For a constant total mass of the disk respect to the maximum grain size (see. Table 2), we derived inner radius of $r_{in}$=${0.27}$ AU, which is comparable for the one obtained by \citet{2009ApJ...692..309E} for the dust sublimation radius. We explored this parameter in the range 0.1-0.4 AU. This parameter mostly affects the SED in the range 1--3 $\mu$m. Increasing $r_{in}$ causes less flux in the range 1--3 $\mu$m and a shift of the maximum flux towards longer wavelengths. Besides, increasing $r_{in}$ causes a decrease of both Keck and MIDI visibilities. But the effect of the increasing $r_{in}$ is more sensible for the Keck visibility than the MIDI ones. At 2 $\mu$m, the warm dust directly defines the inner rim.
             \\

 The range of values which have been explored and the values according to the best model are summarized in Table \ref{table:2}. The more sensitive parameters with respect to the visivilities are ranked first in this Table. 
 
\subsubsection{$ \chi^{2} $ minimisation}

On the basis of our modeling results, multi-wavelength images (from which visibilities are derived in the FOV of the interferometer) and the spectral energy distribution of the simulated disk are computed. The Keck visibilities are calculated for the wavelengths in the range 2–-2.3 $ \mu $m. Since we are focusing on the IR domain, the synthetic SED is compared to the ISO measurements between 2.2 and 50 $ \mu $m.
We calculated reduced $ \chi^{2} $ separately for the SED in the range 2.2--50$ \mu $m, for the Keck visibility (vis1), for the MIDI visibility (vis2) with respect to the free parameters for each model. Then we added all together and divided by 3 to find the reduced total $ \chi^{2} $. In this way, we consider the same weight for the SED , Keck and MIDI visibilities.

\begin{figure}[h]
		\centering
			\includegraphics[width=7cm, height=6cm]{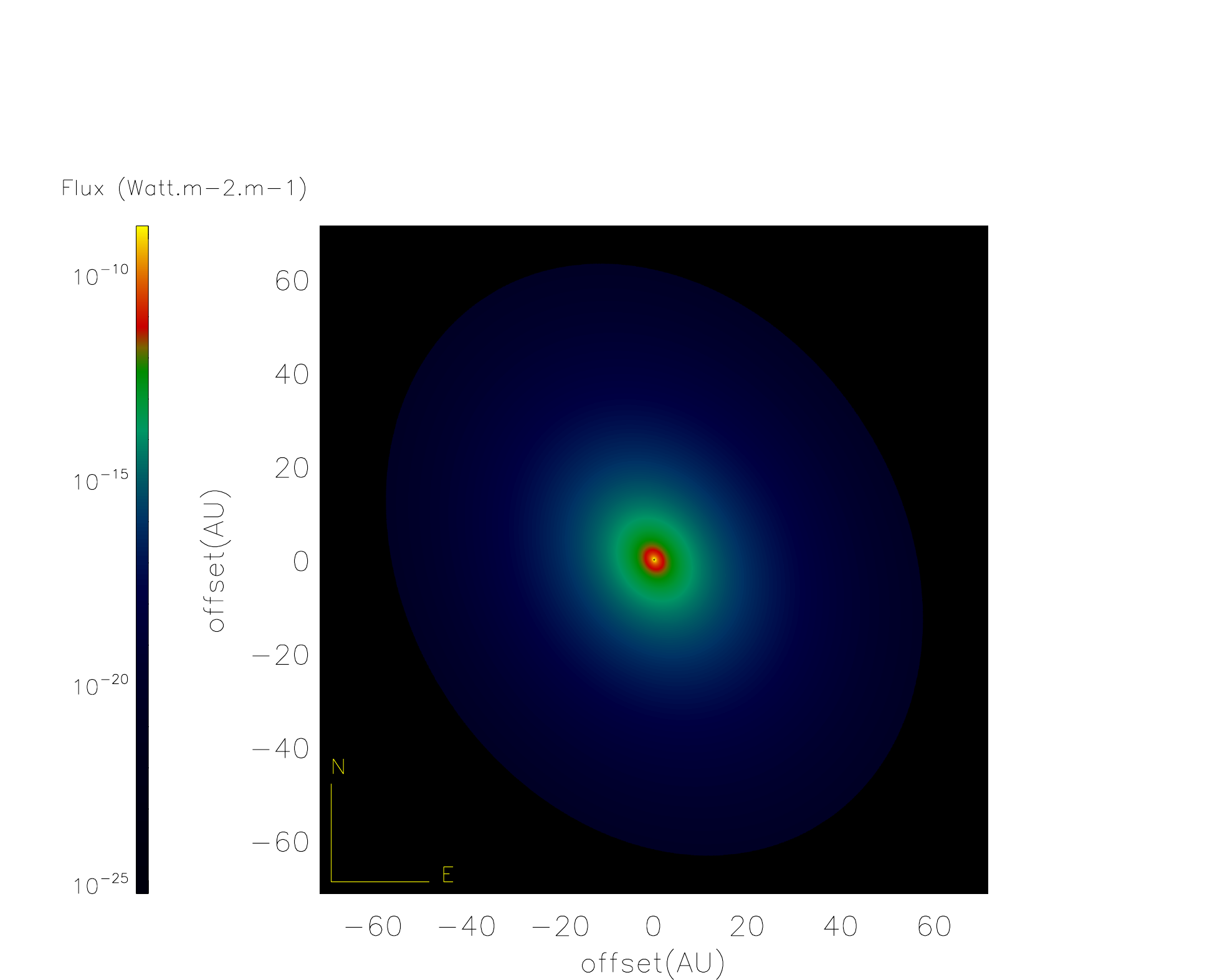}
	\caption{The synthetic image of MWC480 at $\lambda$=10 $\mu$m for the one-component disk model.}
	\label{Fig:1}
\end{figure}
\begin{figure*}[htb]
	\centering
	\includegraphics[width=16cm, height=10cm]{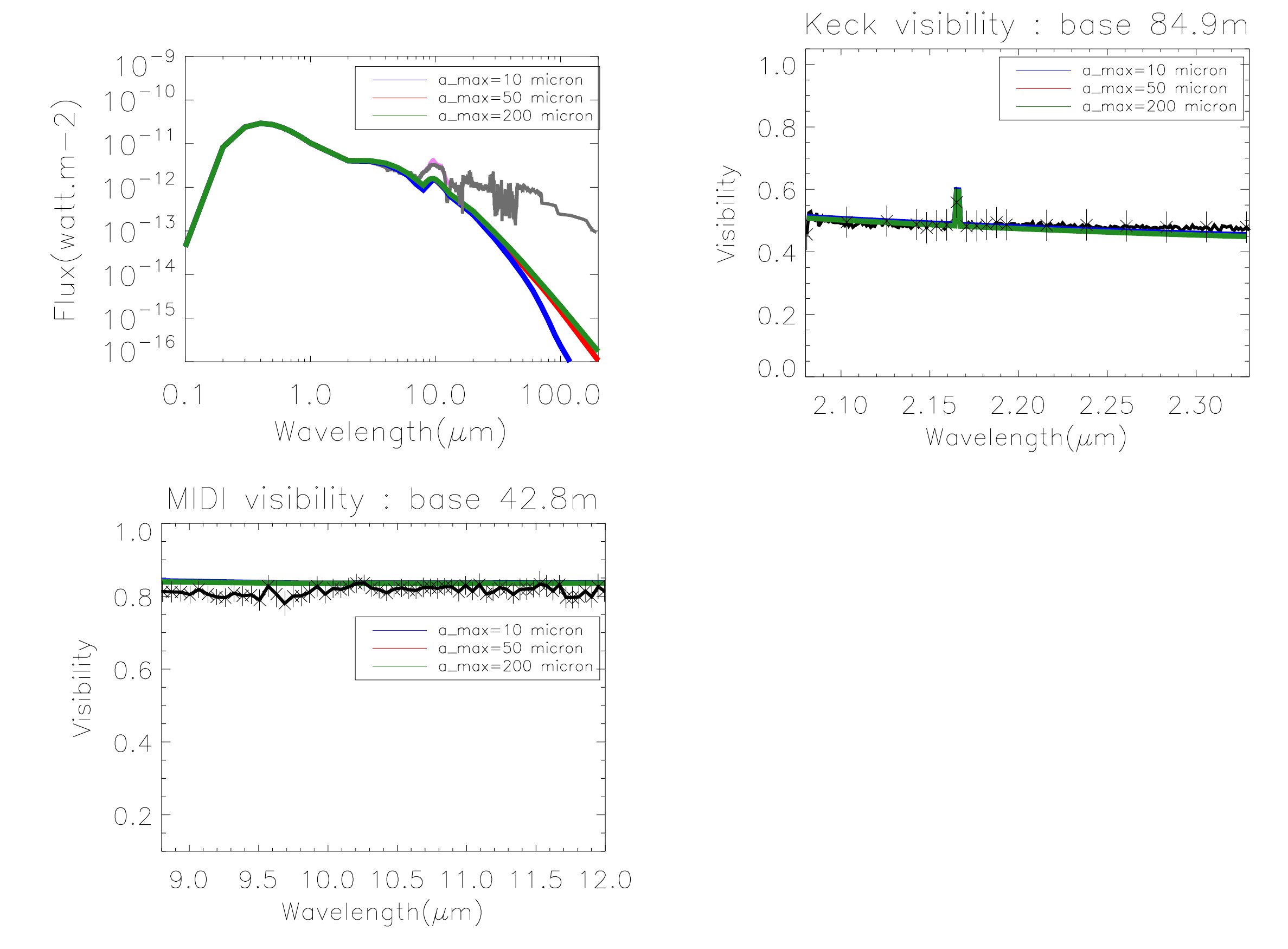}
	\caption{One-component disk model. The effect of the maximum grain size ($a_{max}$ = 10 $\mu$m, $a_{max}$=50 $\mu$m and $a_{max}$= 200 $\mu$m) is compensated by an adjustment of the dust mass:  ${M_{dust}}$=2.4$\times$ ${{10}^{-11}}$ ${M_{\odot}}$, ${M_{dust}}$=0.5$\times$ ${{10}^{-10}}$ ${M_{\odot}}$ and ${M_{dust}}$=${{10}^{-10}}$ ${M_{\odot}}$, respectively. The gray and black solid lines in plots represents the observational data. For the SED, solid gray lines: SpeX and BASS data for 0.8.-14 $\mu$m and ISO data in 1998 for 15--200 $\mu$m. For the SED, pink line: MIDI uncorrelated spectrum between 8--13.5 $\mu$m.}
	\label{Fig:2}%
\end{figure*}

\subsubsection{Results}
The one-component disk model reproduced simultaneously the SED from wavelengths 2 $\mu$m to 8 $\mu$m and the Keck and MIDI visibilities as well. In fact, an optically thin disk decreases the near-IR excess described in \citet{2007ApJ...669.1072E}. {\bf{In Fig. 2}}, we show the synthetic image of this model. In Fig. 3, we represent the SED, the Keck and the MIDI visibilities. According to Fig. 3, the maximum grain size of the dust has a weak influence on the near-IR emission. On the other hand, there is a small effect on the wavelengths 30--70 $\mu$m in the SED. MIDI visibilities does not change for the three maximum grain sizes. The maximum grain size of 50 and 200 $\mu$m are more consistent with the SED (See. Fig.3). 
According to Table 2, the reduced $ \chi^{2} $ for the SED is much larger than the value we obtained for the reduced $ \chi^{2} $ for the visibilities.
 
There are, however, one issue in this model:
         \\
         
 The model presented a deficit of the emission at wavelengths 8-200 $\mu$m. We then may need to use another component for the disk of MWC480 to compensate this lack of longer wavelengths emission.

\begin{table*}[h]
\caption{The best parameters and their explored ranges for the two-components disk model}             
\label{table:3}      
\centering
\begin{tabular}{c|c|c|c}
\hline
\hline
 & \textbf{First component } &  &\\
\hline
\hline
\textbf{Parameters} & \textbf{Best-Values} & \textbf{Explored Ranges} & \\
\hline
\textbf{$M_{dust}$($a_{max}$ = 10 $\mu$m)} & 2.46  $\times$ ${{10}^{-11}}$ ${M_{\odot}}$ & ${10}^{-12}$...${10}^{-7}$ ${M_{\odot}}$\\
\hline
\textbf{$M_{dust}$($a_{max}$ = 50 $\mu$m)} & 0.47  $\times$ ${{10}^{-10}}$ ${M_{\odot}}$ & ${10}^{-12}$...${10}^{-7}$ ${M_{\odot}}$\\
\hline
\textbf{$M_{dust}$($a_{max}$ = 200 $\mu$m)} & 0.9  $\times$ ${{10}^{-10}}$ ${M_{\odot}}$ & ${10}^{-12}$...${10}^{-7}$ ${M_{\odot}}$ \\
\hline
\textbf{p} & ${1.5}$ & 0.1...1.98 \\
\hline
\textbf{q} & ${0.5}$ & 0.4...0.9 \\
\hline
\textbf{$r_{in}$} & 0.27 AU &fixed \\
\hline
\hline

 & \textbf{Second component } &  &  \\
\hline
\textbf{Parameters} & \textbf{Best-Values} & \textbf{Explored Ranges} \\
\hline
\textbf{$M_{dust}$($a_{max}$ = 10 $\mu$m)} & 0.65  $\times$ ${{10}^{-7}}$ ${M_{\odot}}$ & ${10}^{-12}$...${10}^{-7}$ ${M_{\odot}}$& \\
\hline
\textbf{$M_{dust}$($a_{max}$ = 50 $\mu$m)} & 0.75  $\times$ ${{10}^{-7}}$ ${M_{\odot}}$ & ${10}^{-12}$...${10}^{-7}$ ${M_{\odot}}$ &\\
\hline
\textbf{$M_{dust}$($a_{max}$ = 200 $\mu$m)} & 1.25  $\times$ ${{10}^{-7}}$ ${M_{\odot}}$ & ${10}^{-12}$...${10}^{-7}$ ${M_{\odot}}$ & \\
\hline
\textbf{p} & 0.6 & 0.1...1.98 & \\
\hline
\textbf{q} & 0.5 & 0.4...0.9 & \\
\hline
\textbf{$r_{in}$} & 52 AU & 10...60 AU & \\
\hline
\textbf{$r_{out}$} & 80 AU & fixed & \\
\hline
\hline

\textbf{$ \chi_{r \hspace{0.1cm} SED}^{2} $} & \textbf{$ \chi_{r \hspace{0.1cm} vis1}^{2} $} & \textbf{$ \chi_{r \hspace{0.1cm} vis2}^{2} $}&\textbf{$ \chi_{r \hspace{0.1cm} total}^{2} $} \\
\hline
20.0& 0.06& 2.3& 7.45\\
\hline
\end{tabular}
\end{table*}


\begin{figure*}[htb]
	\centering
	\includegraphics[width=16cm, height=10cm]{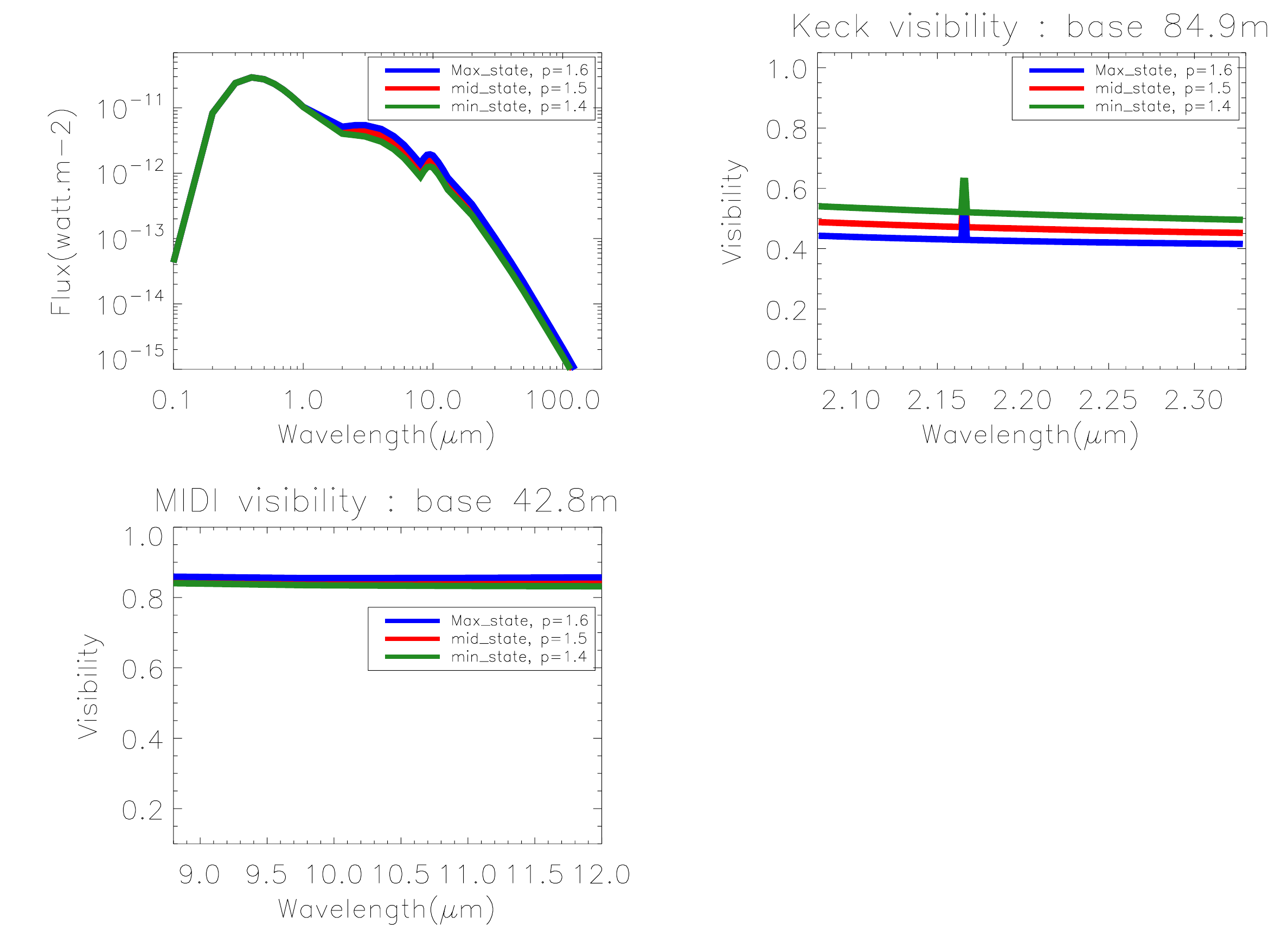}
	\caption{The effect of variability of the star in the near-IR in different brightness state for the maximum grain size of 200 $ \mu $m in one-component disk model .}
	\label{}%
\end{figure*} 

In Fig. 4, we modelized the variabilities assumed to be caused by the scale height of the inner disk. Different brightness state can be produced in the SED by modifying the exponent of the surface density law, p for constant dust mass and inclination. The p value is linked to the change of the scale height in the inner disk. The mid brightness state corresponds to the best parameter we found for p for the one-component disk model, which is almost consistent with the near-IR data in 2007. We do show that the variability of the scale height in the inner disk  affects especially the Keck visibilities. In the maximum illumination state in the near-IR compared to mid and minimum one, there is more flux for the wavelengths from 1 $\mu$m to 13 $\mu$m. Besides, for the Keck visibilities, the maximum state has a lower visibility value compared to the mid and the minimum one, which means that the disk in the near-IR is more resolved for the maximum state. For the MIDI visibilities, the effect of brightness variability in the mid-IR induce a small change for the wavelengths 10--12 $ \mu $m. Indeed, the disk in the mid-IR is more resolved when the source is in its minimum brightness state.

Since variability of the source in the near-IR affects the visibilities, it is very important to use the data in the SED, which are taken in the same time than intereferometric observations.

However, even considering the variability in the near-IR as maximum or minimum illumination state, the lack of emission in the SED in the longer wavelengths ($ \lambda$ $>$ 10-20  $\mu$m ) can not be explained in one-component disk model.




\subsection{Application to the two-components disk model}


 We found that using a second component in the disk of MWC480 is necessary since the one component for the disk of MWC480 could not reproduce the longer wavelengths emission. As we did show in Section 3.1.2, the variability of the star in the near-IR also can not explain the lack of emission in the longer wavelengths. We considered a second component for the disk. We assumed that the first component is optically thin and the outer component is optically thick. The cool dusts in the outer component of the disk emit at longer wavelengths which can compensate the lack of emission at longer wavelength in the SED.

The parameters in this model are:

\begin{enumerate}
\item[1)] The total dust mass $M_{dust}$ for the first component,
\item[2)] The total dust mass $M_{dust}$ for the second component,
\item[3)] the surface density power-law exponent p for the first component,
\item[4)] the surface density power-law exponent p for the second component,
\item[5)] the temperature power-law exponent q for the first component,
\item[6)] the temperature power-law exponent q for the second component,
\item[7)] the inner radius $r_{in}$ for the first component, 
\item[8)] the inner radius $r_{in}$ for the second component, 
\item[9)] the outer radius $r_{out}$ for the second component 
\end{enumerate}
We keep the inclination, P.A. and the outer radius of the disk identical
 to the values we considered 
\begin{figure}[h]
	\centering
	\includegraphics[width=7cm, height=6cm]{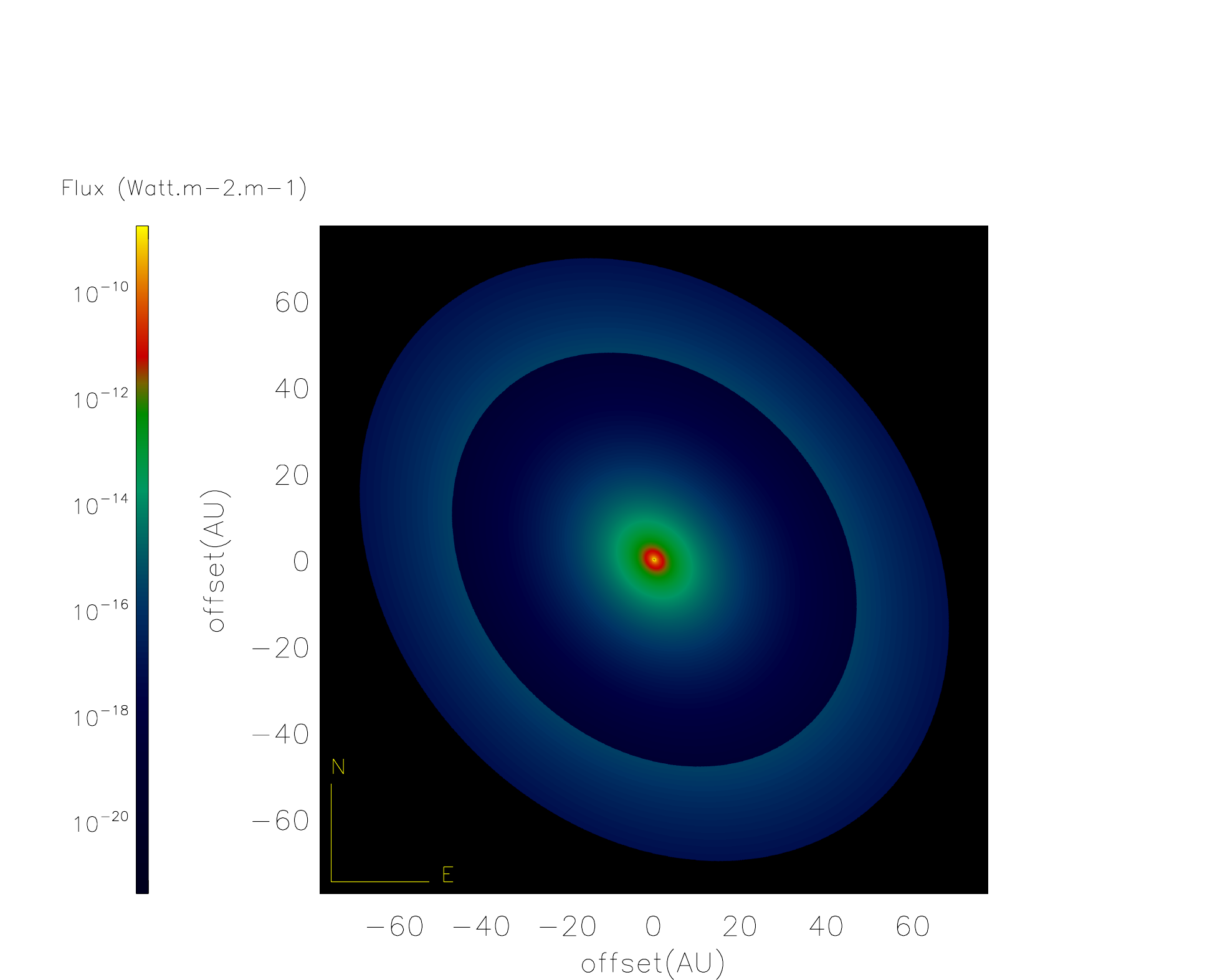}
	\caption{The synthetic image of MWC480 at $\lambda$=10 $\mu$m for the two-components disk model.}
	\label{Fig:5}
\end{figure} 
 for the one-component disk model.

Since the two components of the disk are attached, then the outer radius of the first disk is equal to the inner radius of the second component. 
To minimize our free parameters, we also keep the best value of the inner radius of the disk we found for the one-component disk model $ r_{in}=0.27 AU $, which is comparable for the one obtained by \citet{2009ApJ...692..309E} for the dust sublimation radius.

We explored all the free parameters to find the best model corresponding to our data. In Table \ref{table:3}, we summarize the best fitting parameters.
{\bf{In {Fig. 5}, we show the synthetic image of this model}}. In Fig. 6 we represent the SED, the Keck and the MIDI visibilities for three different maximum grain sizes of the grain size distribution.

\subsubsection{$ \chi^{2} $ minimization}
In this model, we use the same method we used for one-component disk model to find the reduced total $ \chi^{2} $ (See. Section 3.1.1). For computing the reduced $ \chi^{2} $ for the SED, the Keck and the MIDI visibilities, we consider the 7 free parameters of the two-componnets disk model.

In Table 3, we show the reduced $ \chi^{2} $ for the SED, Keck and MIDI visibilities and the total reduced $ \chi^{2} $ for the two-components disk model.

\subsubsection{Results}
The two-components disk model could reproduce the deficit of the emission at wavelengths from 10-200 $\mu$m and the Keck and the 2007 MIDI visibilities. Although, at 10 $\mu$m still we have a minor deficit of the flux compared to the observations. Decreasing the inner radius of the second component less than 52 AU was not consistent with the SED and visibilities. The maximum grain size of 200 $\mu$m of the grain distribution was more consistent with the SED.
                    \\
   
One thing we should emphasize here is that the total dust mass, dominated by the second component, is about 1000 times smaller than the dust mass derived from continuum millimeter measurements. \citet{1997ApJ...490..792M} using millimeter observations, found a dust mass of 2.4$\times$ $10^{-4}$--2.9$\times$ $10^{-4}$ $M _{\odot} $. Whereas the dust disk mass derived in our paper is 1.25$\times$ $10^{-7}$ $M _{\odot} $ only. Studying the star MWC480 in the millimeter wavelengths were done with an observed continuum emitted by an optically thin medium \citep{1997ApJ...490..792M}. By observing an optically thin medium, it is possible to have a direct and real measure of the dust mass. We plot in Fig. 7 the vertical optical depth  at 10 $\mu$m versus distance from central star for the two components disk of star MWC480. According to Fig. 5 , since the optical depth of the first component is less than 1, then the value obtained for the mass of the dust is a real estimate of the total amount of the dust in this component. The optical depth of the second component of the disk is greater than 1 (see Fig. 7 ). Increasing the mass of the dust for the second component more than 1.25$\times$ $10^{-7}$ $M _{\odot} $, which would contribute to the total dust mass, does not change the SED and the visibilities. The best value found for the dust mass of the second component in our model indeed is the minimum mass required for the dust for this component. It is thus possible to have more dust in the second component as well. Therefore, there is a  contradiction between 2.4$\times$ $10^{-4}$ dust mass determined by mm observations and 1.25$\times$ $10^{-7}$ $M _{\odot} $ one, which is the minimum mass determined from mid-IR.
                        \\

We suspect that the disk around MWC480 may be a pre-transitional disk and the inner parts are (almost) depleted of material. Fig.8 shows that there is a discontinuity in the transition between two components disk of MWC480 in the surface density from $ 4\times10^{-7} $ to $ 1.5\times10^{-3} $ kg/$ cm^{3} $. This kind of jump in the surface density between two components is seen for many of pre-transitional disks such as Herbig star HD 100546. 

\begin{figure*}[htb]
	\centering
	\includegraphics[width=16cm, height=10cm]{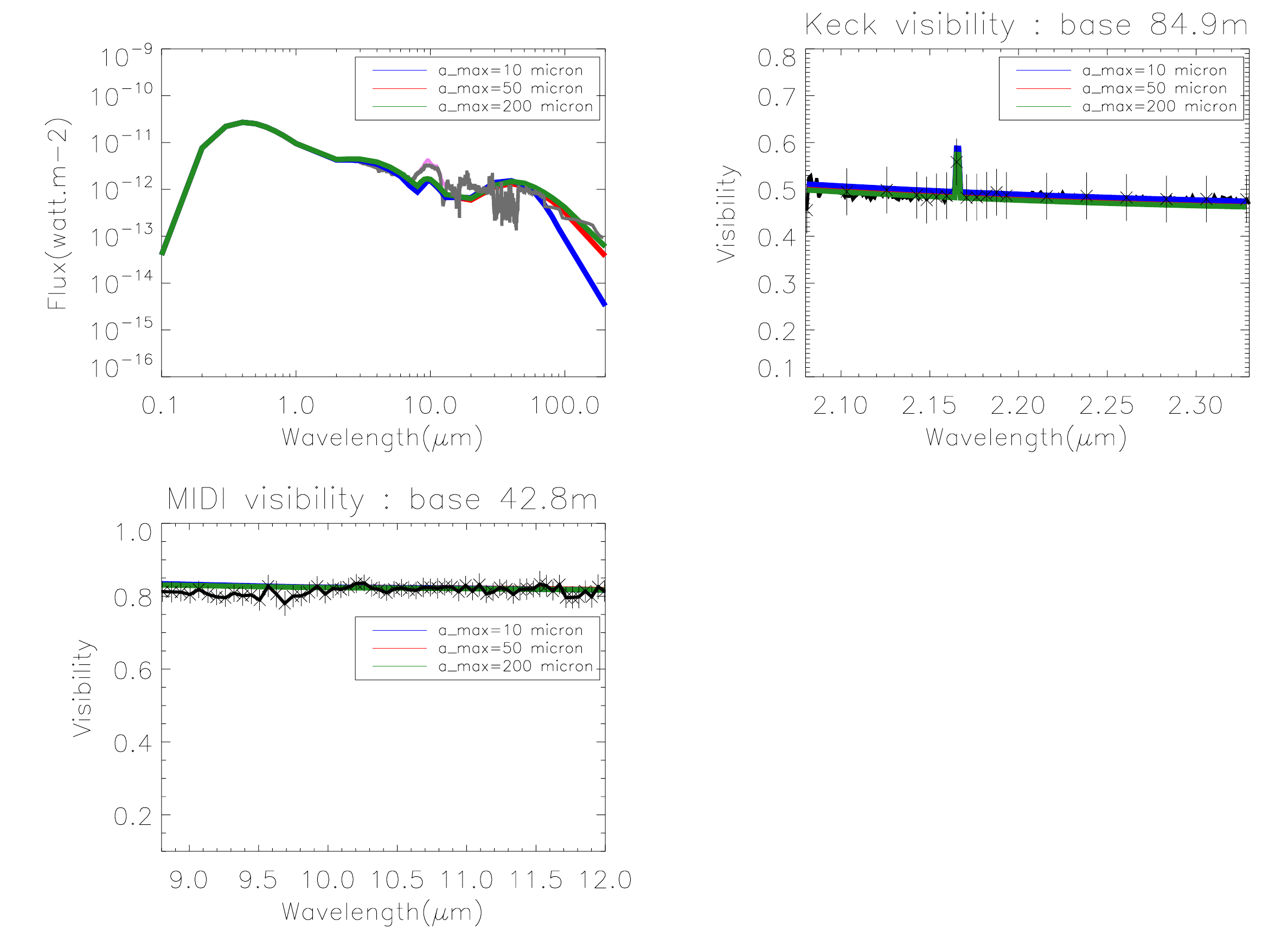}
	\caption{Two-components disk model. The gray and black solid lines in plots represents the observational data. For the SED, solid gray lines: SpeX and BASS data for 0.8.-14 $\mu$m and ISO data in 1998 for 15--200 $\mu$m. For the SED, pink line: MIDI uncorrelated spectrum between 8--13.5 $\mu$m.}
	\label{Fig:4}%
\end{figure*}

\citet{2011A&A...531A...1T} show a surface density jump between two components disk of HD 100546 is an order of  $\sim 10^{-4} $. However, the second component of HD 100546 considering the gap is located at 13 AU and the one MWC480 without gap at 52 AU. According to \citet{2011A&A...531A...1T}, the second component of HD100546 has a mass of $ \sim  10^{-4} $ $M _{\odot} $, which is $ 10^{3} $ times greater than the mass of second component of MWC480, 1.25$\times$ $10^{-7}$ $M _{\odot} $. However, the second component disk of HD 100546 is more extended than the MWC480 one. The position of the second component disk of MWC480 is roughly similar to the pre-transitional disk around LkCa 15 one at 58 AU \citep{2010ApJ...717..441E}. 
         \\
         
In the case of MWC480, although the inner component compared to the second one has significantly lower mass, there is no dip in the infrared emission of the SED. However, \citet{2012ApJ...747..103E} figuring out the status of the disks around 14 stars, showed that the dip in the infrared emission of SED of star LRLL 37 is not obvious and on the other hand a full disk could not reproduce its SED because of strong silicate emission seen in this object. So they conclude that this could be a sign that LRLL 37 is a pre-transitional disk with a 
\begin{figure}[h]
	\centering
	\includegraphics[width=6cm, height=4cm]{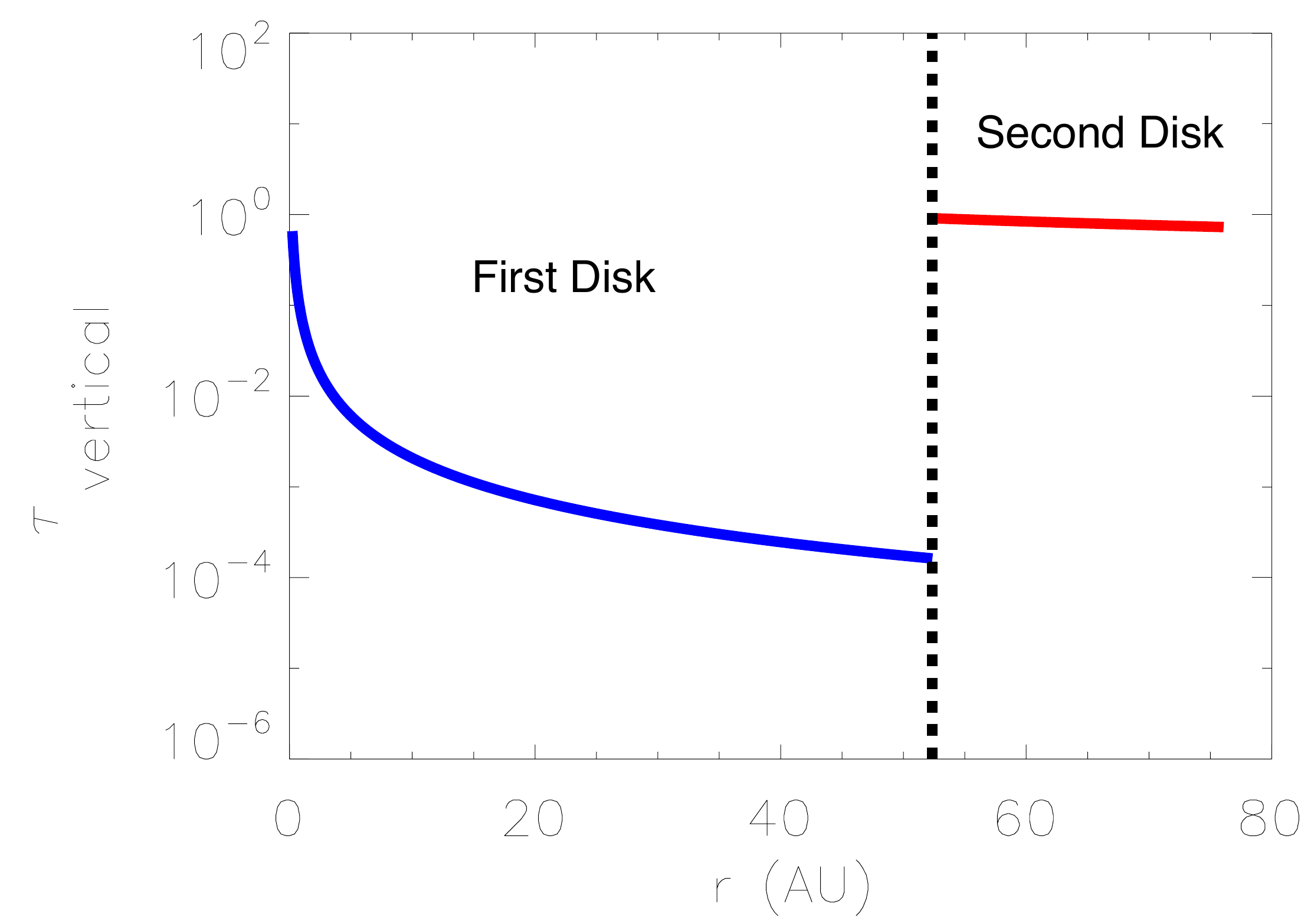}
	\caption{The vertical optical depth at wavelengths 10 $ \mu $m versus the radius for the two-components disk model. The dashed line shows the transition between the two components.}
	\label{Fig:2}%
\end{figure}

small gap that contains some small optically thin dust in the inner component. We may assume that in the inner component the dust dissipation has begun and thus MWC480 is a pre-transitional disk source as LRLL 37.

According to Table 3, the reduced $ \chi^{2} $ for the SED is larger than the reduced one for the visibilities. 
In fact, considering the reduced $ \chi^{2} $ for the visibilities (See. Table 2) the one-component disk model can be consistent with the interferometric data. 

\begin{figure}[h]
	\centering
\includegraphics[width=6cm, height=4cm]{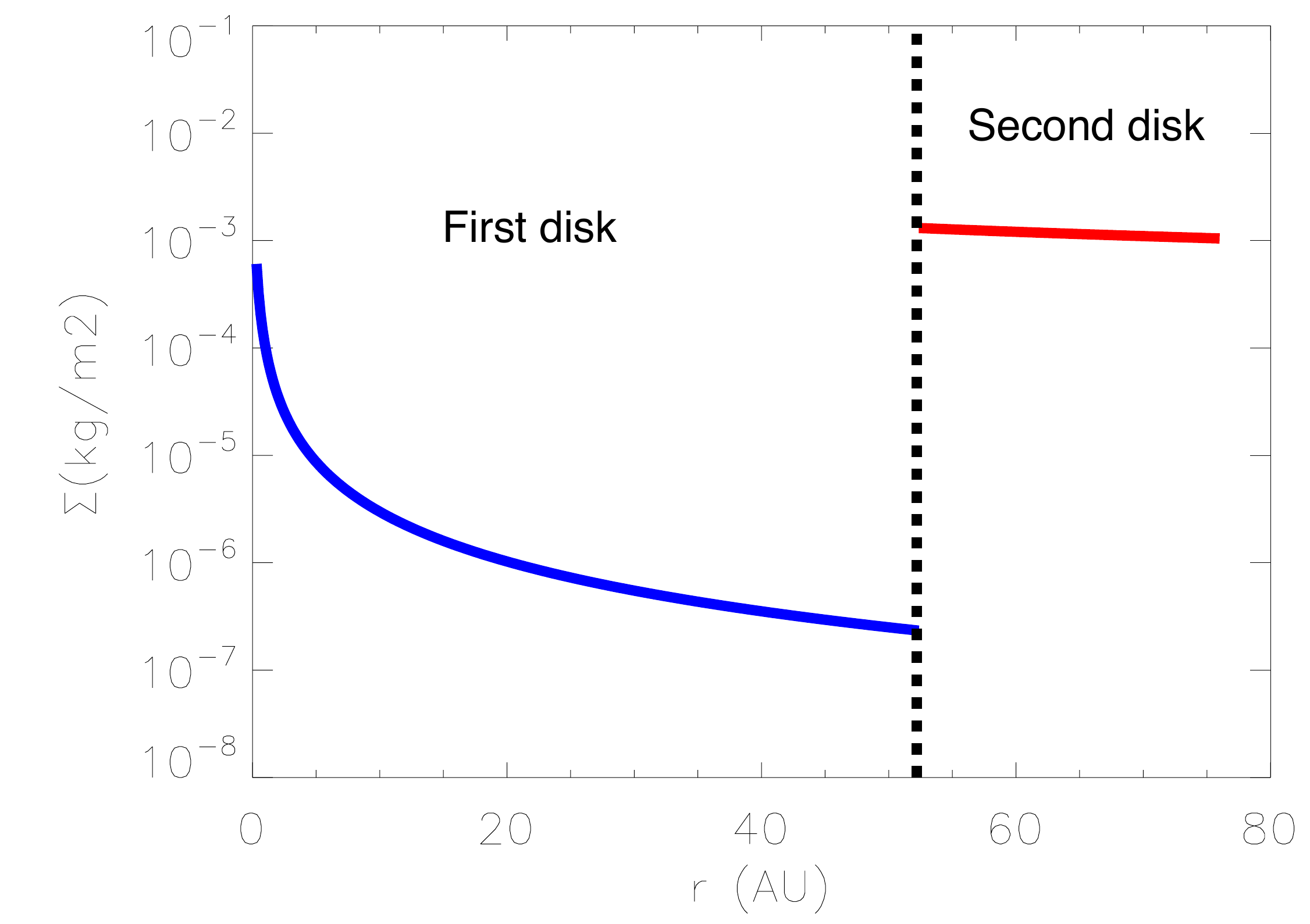}
	\caption{The surface density at wavelength 10 $ \mu $m versus the radius for the two-components disk model. The dashed line shows the transition between the two components.}
	\label{Fig:2}%
\end{figure} 
The only case that makes difference between the two models is the SED especially at longer wavelengths.

\section{Summary and perspectives}

 Using observations based on stellar interferometry in the mid-IR, we were able to resolve the circumstellar emission around the Herbig star MWC480. We performed a coherent modeling reproducing the SED and the visibility simultaneously. The modeling is based on a semi-analytical approach using a temperature and surface density-gradient laws. Our aim is to constrain the overall spatial structure of the inner disk region and explore the possible multi-component structure of the disk to better understand the conditions of planet formation in the inner region. In addition, we tried to use three different maximum grain sizes in a size distribution assumed to be of interstellar dust comprising silicate, probably amorphous and graphite composition \citep{1993ApJ...402..441L}. The maximum grain size of 200 $ \mu $m assumed in the distribution was more consistent with the SED of the star. We conclude that:

   \begin{enumerate}
      \item[-] A two-components disk model could reproduce better the SED, the Keck and the MIDI visibilities simultaneously. In one component disk model, increasing the near-IR emission by exploring all free parameters until maximum brightness state even was not consistent with the SED in the longer wavelengths. In fact, the second component of the disk mostly contributes to the emission at longer wavelengths. Then, the second component disk in our modeling is necessary to reproduce all our measurements simultaneously.
   
       \item[-] We suspect that the disk around MWC480 may be a pre-transitional disk since the inner parts are (almost) depleted of the material. The surface density distribution of MWC480 shows a jump in the transition between two components disk, which is seen between components disk of many pre-transitional and transitional disks around Herbig stars. In the SED in the mid-IR wavelengths of MWC480 there is no significant dip. However, the dip in the infrared emission of e.g., star LRLL 37 is not obvious, although the disk around this star is considered as a pre-transitional disk.
      \item[-] Many authors showed that MWC480 presents a time variability in the SED in the near-IR and mid-IR wavelengths. It is now well-established that the near-IR and mid- to far-IR are often anti-correlated, at least in transitional disks. The most likely scenario is changes in the scale height of the inner disk, which emits in the near-IR wavelengths. This leads to changes in the shadowing of the outer disk so that the illumination by the central star changes with time. This affects both the scattered light and thermal emission of the entire disk. 
      In this paper, first, we show that the total flux of the MIDI observations in 2007 is almost consistent with the BASS data in 2007 in N band. Figuring out the effect of time variability of the star MWC480 in the near-IR in our models, we found that the maximum or minimum brightness state affects on the visibilities especially the Keck ones. The effect on the SED is significant only for wavelengths 2--20 $ \mu $m. For the maximum brightness state in the near-IR, the value of the Keck visibilities is less than for the minimum state. For MIDI visibilities, this variability in brightness in the near-IR is less significant. However, for the minimum brightness state, the disk in the mid-IR is more resolved than for the minimum state one.
      Since variability of the star in the near-IR affects on the visibilities, it is very important to use the data, which are taken in the same time of intereferometric observations.
      
      \item[-] We are far from being able to claim that we fully understand the inner disk of star MWC480. For a better understanding of its system, it is crucial that repeated observations at near and mid-IR wavelengths. The MIDI observations have been obtained with only one configurations and one orientation. As said above, a two-components disk model is necessary, however more constraints are required. 
      \item[-] In the future, we would like to image this star with the second-generation VLTI instrument MATISSE (the Multi AperTure mid-Infrared SpectroScopic Experiment) to further assess the inner region of the disk around MWC480. MATISSE will give access to the L, M and N bands. It will be the first time the L and M bands could be used for an interferometric instrument. This two bands give information about hot dusts in the inner region of the disk. With MATISSE, we can increase the number of measurements using 4 telescopes, making a good U-V coverage with different baseline orientations and provide the image reconstruction of the real object with access to the closure phase. The closure phase is an observable quantity, which can be used to reveal the amount of asymmetry in the brightness distribution.  
   \end{enumerate}

\acknowledgments
The MIDI observation of this work has been obtained by Di Folco in 2007. Authors want to thank A. Matter and A. Meilland for useful exchanges.


\bibliographystyle{spr-mp-nameyear-cnd}

\bibliography{biblio-u1}

\end{document}